# Semiconductor nanowire metamaterial for broadband near-unity absorption

[†]Burak Tekcan[1,2], [†]Brad van Kasteren[1,2], [†]Sasan V. Grayli[1,2], Daozhi Shen[1,4,5], Man Chun Tam[2,3], Dayan Ban[2,3], Zbigniew Wasilewski[2,3], Adam W. Tsen[1,3,4,6], Michael E. Reimer[1,2,6]*

[1]Institute for Quantum Computing, University of Waterloo, Ontario, Canada. [2]Department of Electrical & Computer Engineering, University of Waterloo, Ontario, Canada. [3]Waterloo Institute for Nanotechnology, University of Waterloo, Ontario, Canada. [4]Department of Chemistry, University of Waterloo, Ontario, Canada. [5]Centre for Advanced Materials Joining, University of Waterloo, Ontario, Canada, [6]Department of Physics and Astronomy, University of Waterloo, Ontario, Canada.
†These authors contributed equally to this work.

**Abstract**

**The realization of a semiconductor near-unity absorber in the infrared will provide new capabilities to transform applications in sensing, health, imaging, and quantum information science, especially where portability is required. Typically, commercially available portable single-photon detectors in the infrared are made from bulk semiconductors and have efficiencies well below unity. Here, we design a novel semiconductor nanowire metamaterial, and show that by carefully arranging an InGaAs nanowire array and by controlling their shape, we demonstrate near-unity absorption efficiency at room temperature. We experimentally show an average measured efficiency of 93% (simulated average efficiency of 97%) over an unprecedented wavelength range from 900 nm to 1500 nm. We further show that the near-unity absorption results from the collective response of the nanowire metamaterial, originating from both coupling into leaky resonant waveguide and transverse modes. These coupling mechanisms cause light to be absorbed directly from the top and indirectly as light scatters from one nanowire to neighbouring ones. This work leads to the possible development of a new generation of quantum detectors with unprecedented broadband near-unity absorption in the infrared, while operating near room temperature for a wider range of applications.**

The field of optical metamaterials demonstrates promise because of the inherent advantages they exhibit over their bulk counterpart for enhanced absorption or realizing an exotic optical response, which originates from the specific arrangement and the collective interactions of the so-called metaatoms[1–7]. A metaatom is considered the unit cell of a metamaterial, comprised of a single or many structures, whereby their unique arrangement leads to an optical response different from the individual structure. Their coordinated design and careful placement can lead to a near-unity absorber for a broader range of applications. It has been shown that the properties of propagating photons, such as polarization and refraction, through interaction with metamaterials can be exploited to create super-lenses, negative refraction, asymmetric transmission, and cloaking[1–10]. Early demonstrations of such metamaterial properties were achieved by exploiting plasmons (i.e., the collective oscillation of electrons in metals) to control the interaction of light with the surface[1,2,10–12]. In this class of plasmonic metamaterials, the electric dipole response is controlled by the shape, size, and orientation of the metaatoms in the crystalline structure[10–14]. More recently, dielectric and semiconductor metamaterials were used to demonstrate light manipulation with lower losses than their plasmonic counterpart[1,2,15–18]. Indeed, the freedom to control the magnetic and electric field response in such structures leads to the realization of highly absorbing or transmissive metamaterials with unique qualities that are highly desired in sensing and imaging applications[3,5,8,9,15,19–23].

Near-unity absorption in semiconductor metamaterials has been previously demonstrated by overlapping both electric- and magnetic-field components in the photonic nanostructure with Mie resonators and nanowires; however, this near-unity absorption efficiency could only be achieved over a narrow bandwidth in the range of tens of nanometers as reported in previous works[9,24,25]. In this work, we overcome this narrow bandwidth limitation by designing a new semiconductor metamaterial made of indium gallium arsenide (InGaAs) nanowire metaatoms and precisely arranging them into an array. By optimizing the nanowire array geometry and shape, we design a near-unity absorber over an unprecedented wavelength range, from 400 nm to 1650 nm, with an average calculated absorptance of 92%. We fabricated the optimized metamaterial design and confirmed its infrared performance by Fourier-transform infrared (FTIR) spectroscopy. The metamaterial exhibited near-unity absorption over the wavelength range from 900 nm to 1500 nm, with a measured average absorption efficiency of 93% (97% calculated). We compare this measured absorption of the InGaAs nanowire metamaterial to bulk InGaAs of similar thickness and show that the absorptance is significantly enhanced by the nanowire shape and array geometry towards unity. This new class of semiconductor nanowire metamaterials meets the need for a near-unity absorber in the wavelength range where the quantum efficiencies of commercially available detectors are limited without the need for cryogenic cooling. The ability to realize a near-unity absorber in a semiconductor material over such a broad range of wavelengths is of particular interest for developing a next generation single-photon detector for a wider range of applications. Reaching impacts range from quantum technologies in imaging, sensing, communication and computing to biomedical applications such as dose monitoring for cancer treatment and imaging of the eye to identify potentially blinding diseases.

**Mechanisms for Enhanced Absorption in Nanowire Metamaterials**

To design a broadband, near-unity absorber, we first investigate the role of periodicity on the absorption profile in cylindrical nanowires. Although high absorption in cylindrical nanowires has been previously investigated[26–28], the role of periodicity on achieving near-unity absorption has been largely unexplored until now. Figure 1a shows a schematic view of a cylindrical nanowire metamaterial, depicting how leaky resonant modes are being excited in the nanowire array. There are two main avenues for absorption in cylindrical nanowire metamaterials when it is illuminated from the top. First, light can be directly absorbed in the nanowire through coupling to leaky hybrid electric and magnetic modes ($HE_{11}$ and $EH_{11}$)[24,26,27]. Second, light can be indirectly absorbed by scattering from one nanowire to another one through excitation of transverse electric (TE) and transverse magnetic (TM) modes, thus, leading to further absorption. In this latter case, the leaky guided modes migrate towards the nanowire sidewall as they travel downwards. At the interface of the nanowire sidewall and the surrounding dielectric medium, the leaky modes decouple from the nanowire and subsequently emit into free-space. This emitted light then interacts with neighbouring nanowires within the array by coupling to both TE and TM modes. The strong interaction of the electric field between neighboring nanowires for metaatoms with diameter of 200 nm, periodicity of 833 nm and height of 1400 nm is illustrated in Figure 1b. The resulting spatial absorption profile from these mechanisms in the nanowire metaatoms is shown in Figure 1c.

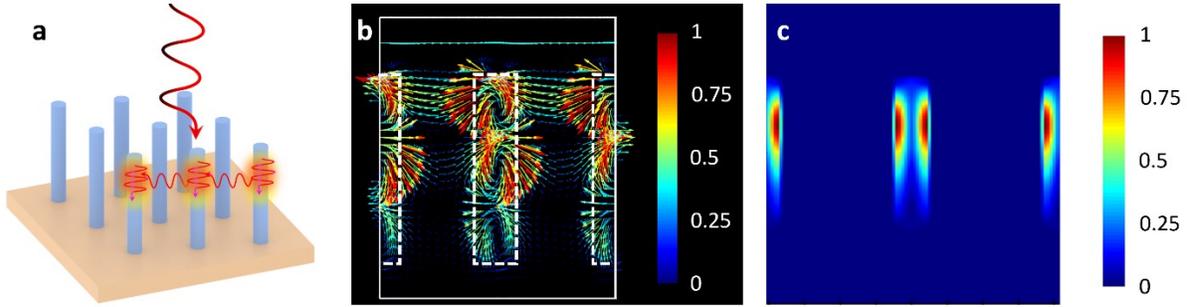

**Figure 1. Mechanisms for enhanced absorption.** (a) Schematic view of the collective interaction of a cylindrical nanowire metamaterial depicting the direct and indirect coupling to leaky guided and transverse modes. The leaky nature of the guided modes in the nanowires leads to free-space scattering of the electric field, which is then recoupled into the neighbouring nanowire metaatoms via excitation of TE and TM modes. (b) Simulated electric field response of a nanowire array with periodicity of 833 nm, nanowire diameter of 200 nm and height of 1400 nm. The interaction of the electric field between neighbouring nanowires is depicted by the vectors and their color indicates the field strength. c) Localized absorption profile (indicated by the color) of the nanowire metamaterial at λ=1020 nm with the same dimensions as (b). In the b) and c) color bars, red represents the maximum magnitude (normalized to 1) and blue is the minimum.

To further understand and explain these two routes of absorption, we investigate the role of how light is coupled into the nanowires as a function of incident angle. Modal analysis for this coupling can be performed using the eigenvalue equation, which satisfies Maxwell's equation, for cylindrical boundary conditions of a wire with radius, $a$, and infinite length[29]:

$$\left(\frac{1}{k_1^2 - k_2^2}\right)^2 \left(\frac{\beta m}{a}\right)^2 = k_0^2 \left(n^2 \frac{J'_m(k_1 a)}{k_1 J_m(k_1 a)} - n_0^2 \frac{H'_m(k_2 a)}{k_2 H_m(k_2 a)}\right) \times \left(\frac{J'_m(k_1 a)}{k_1 J_m(k_1 a)} - \frac{H'_m(k_2 a)}{k_2 H_m(k_2 a)}\right) \quad (1),$$

where $k_1$ and $k_2$ represent the transverse components of the wave vectors (inside and outside of the nanowire, respectively), $n$ and $n_0$ are the refractive indices (inside and outside of the nanowire, respectively), $\beta$ and $k_0$ are wave vectors along the nanowire axis and in free-space, respectively, and $J_m$ and $H_m$ are the $m$th-order of the Bessel function and the Hankel function of the first kind, respectively. The prime indicates the derivative of these Bessel and Hankel functions. The transverse components in equation (1) are angle dependent, where $k_1 = k_0\sqrt{n^2 - cos^2\theta}$, $k_2 = k_0 \sin\theta$, and $\beta = k_0 \cos\theta$. Thus, changes in the angle of incidence govern which modes are excited in the neighbouring nanowires. There are two main incident angles that we consider, perpendicular (θ = 90°) and parallel (θ = 0°) to the nanowire axis. First, with an incident angle corresponding to the wave vector being perpendicular to the nanowire axis, $\beta$ becomes 0. This results in the excitation of purely TE or TM leaky resonant modes. Alternatively, deviation of the wave vector angle from 90° results in the excitation of HE and EH leaky guided modes.

**Narrowband Near-Unity Absorber — Cylindrical Nanowire Metamaterials**

We further study the cylindrical nanowire case by analyzing the simulated absorptance of a single InGaAs metaatom as a function of its radius (Figure 2a). For nanowire radii smaller than 175 nm, the absorption profile is narrowband as the nanowire supports a single resonant leaky guided mode only. As the cylindrical nanowire radius increases in this single-mode regime (<175 nm), the supported resonant leaky modes shift to longer wavelengths. This observation is consistent with what has been previously reported[27]. However, increasing the radius further results in multiple leaky modes to be supported. In this regime of large nanowire radius, multiple leaky modes overlap, and the spectral selectivity vanishes. Thus, the absorption profile broadens as a result. We note that for a larger nanowire radius (>660nm), the absorption profile is very similar to the bulk response of the same material. Although the spectral sensitivity in larger nanowire radii vanishes, the peak absorption efficiency of a single nanowire still remains far from unity.

In order to reach near-unity absorption efficiency we place the single nanowire in an array and optimize the periodicity. The role of the array periodicity on the absorption efficiency of a nanowire metamaterial has not yet been fully explored until now. Our analysis shows that the absorption efficiency of the cylindrical semiconductor nanowire metamaterial is directly linked to the array periodicity. The results of our analysis are summarized in Figure 2b and Figure 2c. Figure 2b shows the calculated absorptance for varying nanowire periodicity and incident wavelength, while Figure 2c takes two-dimensional slices of this three-dimensional plot from Figure 2b. Here, we find the maximum absorptance with the narrowest bandwidth occurs at the nanowire diameter to periodicity ratio of 0.24. Remarkably, by maintaining this ratio for varying lattice constant (periodicity), near-unity absorption efficiency (>99%) can be selectively achieved in the wavelength range from approximately 1000 nm to 1400 nm (see Figure 2c). Furthermore, minimal roll-off in the absorption efficiency as the incident wavelength approaches the InGaAs bandgap is observed (~96% at 1500 nm and ~80% at 1650 nm).

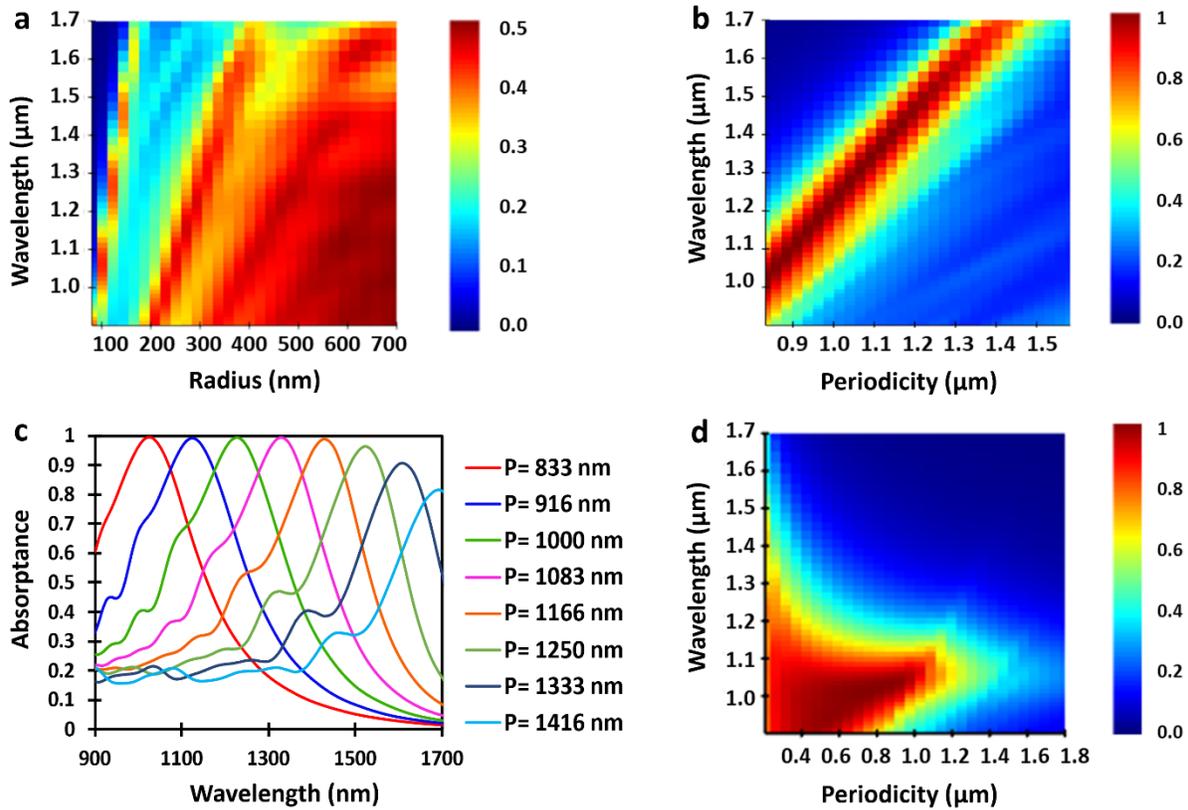

**Figure 2. Finite-difference time-domain (FDTD) simulations of the narrowband absorption efficiency for cylindrical InGaAs nanowires.** (a) Absorption efficiency (color bar) of a single cylindrical nanowire as a function of wavelength and radius. (b) Shifting narrowband near-unity absorption efficiency (color bar) response of the nanowire metamaterial as a function of periodicity with a fixed diameter to lattice constant ratio of 0.24. (c) Two-dimensional slices of the absorption efficiency from (b) showing tunability of the resonant response for near-unity absorption. (d) Absorption efficiency (color bar) of a constant 200 nm diameter cylindrical nanowire metamaterial as a function of array periodicity. The nanowire height used in these simulations from a) to d) is 1400 nm.

To better understand the underlying mechanisms contributing to near-unity absorption in the nanowire metamaterial, we focus our attention on nanowires with dimensions: 200 nm diameter and 1400 nm height. A nanowire height of 1400 nm was selected since this was close to the smallest height where broadband near-unity absorptance was achieved (see Supplementary Note 4). The calculated absorption efficiency with these nanowire dimensions as a function of periodicity is presented in Figure 2d. Near-unity absorption (>99%) occurs for a range of periodicities between 676 nm to 885 nm. In this range of nanowire periodicities, the process of indirect absorption attributed to the decoupling of leaky resonant guided modes and excitation of TE or TM modes in neighbouring nanowires is enhanced. This enhancement maximizes the overall absorption efficiency and highlights its dependence on the nanowire periodicity. Interestingly, this process of near-unity absorption only occurs when the nanowire spacing is optimized. When deviating from this optimized spacing the spectral sensitivity vanishes and the absorption efficiency drops. A smaller nanowire periodicity leads to a broadened absorption spectrum with a lower maximum. In contrast, the absorption spectrum remains narrow for a larger nanowire periodicity, but the maximum gradually drops as the interaction between neighbouring nanowires diminishes.

**Broadband Near-Unity Absorber — Tapered Nanowire Metamaterial**

While cylindrical nanowire metamaterials offer narrow bandwidth, near-unity absorption capabilities, certain sensing applications require broad spectral range detection. One successful approach with a nanowire metamaterial utilizes tapered nanowire metaatoms to accommodate the coupling of a broad range of frequencies[26,30–33]. In this strategy, the nanowire tapering provides a continuum of diameters for all wavelengths to couple to resonant HE and EH leaky guided modes. This broadband behaviour combined with a proper choice of semiconductor material can produce a desirable broadband metamaterial absorber in the infrared.

To design a near-unity broadband absorber in the infrared, we first optimize the tapered InGaAs nanowire shape of a single nanowire by varying the bottom radius for a fixed top radius. The top radius was selected (r = 170nm) to coincide with the cylindrical nanowire radius that produced the highest resonant spectral response near the edge of the InGaAs bandgap (1650 nm). The bottom radius of the tapered nanowire was increased in size until the absorption was enhanced over a broad spectral range (see Figure 3a and Figure 3b). The optimal combination of parameters that we found included a top nanowire radius of 170 nm, bottom nanowire radius of 440 nm and height of 1400 nm.

Next, the optimal nanowire metaatom from Figure 3a was placed in a two-dimensional square lattice to form a metamaterial where the lattice constant can be leveraged to achieve broadband, near-unity absorption. In Figure 3c, we plot the calculated absorption efficiency as a function of periodicity. For this metaatom, we determine that broadband, near-unity absorption in the metamaterial is achieved at a lattice spacing of 900 nm. Increasing the periodicity leads to a decrease in the absorption efficiency as the interaction between neighbouring nanowires weakens. Similar to cylindrical nanowire metamaterials, this behaviour implies that the larger lattice constant reduces the effective recollection of the scattered field by the tapered nanowire array. We note that a lower bound of 900 nm was set for the periodicity since a smaller separation would result in the nanowires overlapping, reducing their selected height. A more detailed description of the nanowire metamaterial optimization process is presented in Supplementary Notes 1-5.

In Figure 3d, we plot the calculated absorption efficiency over an extended wavelength range (400 nm to 1650 nm) at the optimal periodicity to highlight near-unity absorption over an unprecedented bandwidth. The calculated average absorption efficiency is found to be 92%, with peak efficiencies of 99% at 909 nm and 98% at 1406 nm. We note that the observed peaks in Figure 3d result from the nanowires being suspended in air; however, these peaks are later found to be less prominent in our updated model after including a substrate to better compare with the experimental results.

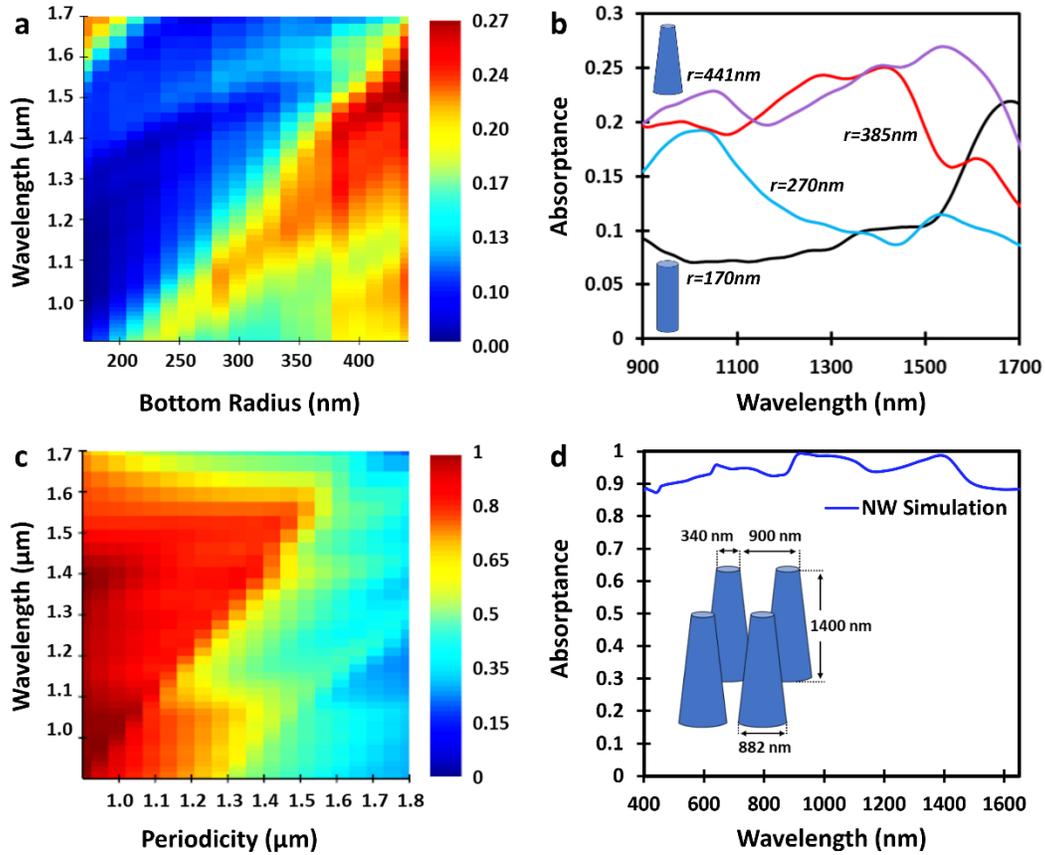

**Figure 3. Finite-difference time-domain (FDTD) simulations of the broadband absorption efficiency for tapered InGaAs nanowires.** (a) Calculated absorption efficiency (color bar) of a single tapered nanowire as a function of bottom nanowire radius with a fixed top radius of 170 nm and height of 1400 nm. (b) Two-dimensional slices of the absorption efficiency from (a) of a single nanowire with varying bottom radius (170 nm, 270 nm, 385 nm, 441 nm) and a fixed top radius of 170 nm. The nanowire is cylindrical at r = 170 nm as indicated by the bottom inset. Increasing the bottom radius leads to a nanowire tapering as indicated by the top inset for r = 441 nm. (c) Calculated absorption efficiency (color bar) dependence on the nanowire periodicity for a nanowire height of 1400 nm, top radius of 170 nm and bottom radius of 441 nm. (d) Calculated absorption efficiency over an extended wavelength range for an optimized nanowire metamaterial, demonstrating near-unity absorption over an unprecedented wavelength range from 400 nm to 1650 nm. The optimized dimensions (shown in the inset) were found to be: lattice constant: 900 nm; bottom diameter: 882 nm; top diameter: 340 nm; and height: 1400 nm.

**Results**

Next, we fabricate the nanowire metamaterial absorber with dimensions as close as possible to the optimized structure using an InGaAs film (2250 nm thickness) grown by molecular beam epitaxy on an (100) InP substrate. We utilized a multistep nanofabrication process to realize a 100 μm x 100 μm metamaterial comprised of a tapered InGaAs nanowire array. See Supplementary Note 6 for details on the nanowire fabrication process.

A scanning electron micrograph of the fabricated optimized nanowire metatoms is shown in the inset of Figure 4a. This image illustrates the uniformity of the tapered nanowires that was achieved during the fabrication process. The high absorption efficiency of the nanowire metamaterial is apparent in the optical image of Figure 4a as the array appears black when compared to the blue-grey InGaAs film. We performed FTIR spectroscopy to measure the absorption efficiency of the nanowire array. A tungsten light source was used to illuminate the sample, and a liquid-nitrogen-cooled Mercury-Cadmium-Telluride (MCT) detector was employed to measure the intensity of the reflected and transmitted spectra. To achieve a spot size smaller than the nanowire metamaterial dimensions in these measurements, a Schwarzschild reflective objective with a numerical aperture of 0.5 was used. The measured reflectance as a function of wavelength is shown in Figure 4b, illustrating the low reflection from the nanowire metamaterial. The reflectance is found to be below 4% over the entire wavelength range from 900 nm to 1650 nm, with an average reflectance of 3%. These results are compared to simulations of the same nanowire metamaterial dimensions, which show excellent quantitative agreement. We note that the fabricated nanowire metamaterial dimensions closely resemble the optimized structure from the numerical calculations. The fabricated nanowire metamaterial (modelled metamaterial) was found to approximately have a top radius of 175 nm (170 nm), bottom radius of 440 nm (441 nm), height of 1300 nm (1400 nm), and a pitch of 900 nm (900 nm). We therefore updated the dimensions in our model for a direct comparison with experiment. We also note that the reflective objective in the FTIR experimental setup introduces a range of angles for light incident on the nanowire array from normal (0°) to 30°. Additionally, the nanowire array is fabricated from an InGaAs film on an InP substrate; however, up to this point the numerical models only considered InGaAs nanowires suspended in air. Thus, these factors were accounted for in our improved model for a more realistic comparison (further details in Supplementary Notes 3 and 4).

We now bring attention to the measured broadband near-unity absorption, as shown in Figure 4c. Remarkably, we measure an average absorptance of 93% from 900 nm to 1500 nm, with a peak efficiency of 95%. Comparing these experimental results with the simulated data in Figure 4c, an excellent agreement is attained with minimal discrepancy. The small deviation that is observed between simulation and experiment may be due to the differing sidewall geometries of the fabricated and modelled nanowires. To obtain the absorption efficiency from the measured results, we utilize A = 1 – R – T, where A, R and T are the absorptance, reflectance, and transmittance, respectively. In our measurement, we found that the transmission data also comprises of thin film interference due to the presence of the reflective objective. We therefore corrected for this interference in the measured absorptance. Details of this correction procedure is presented in Supplementary Note 8. We note that the process we implemented to correct for the thin-film interference underestimates the measured absorptance, thus providing a conservative lower bound to the absorption efficiency of the nanowire metamaterial.

To demonstrate the enhanced absorption in tapered nanowire metamaterials we compared it to an unprocessed bulk InGaAs film with a thickness of 2250 nm. The measured bulk InGaAs film absorptance is shown as a function of wavelength in Figure 4d (red solid curve). This experimental data is compared to the simulated bulk InGaAs film with an identical thickness of 2250 nm and demonstrates excellent agreement (black dashed cross). The enhancement of the absorption towards unity when utilizing the tapered nanowire geometry is evident when comparing tapered nanowires (Figure 4c) and the planar bulk material (Figure 4d). We also simulated bulk InGaAs with an identical thickness (1300 nm) to the thickness of the tapered nanowire array (green dashed circle) and a thickness (1000 nm) of commercially available InGaAs photodetectors for reference (blue dashed diamond). Numerical calculations and the measured performance for a series of varied nanowire geometries were also studied (see Supplementary Note 10). Despite deviating from the optimized structure, our measurements show the reproducibility of enhanced absorption in tapered nanowire metamaterials.

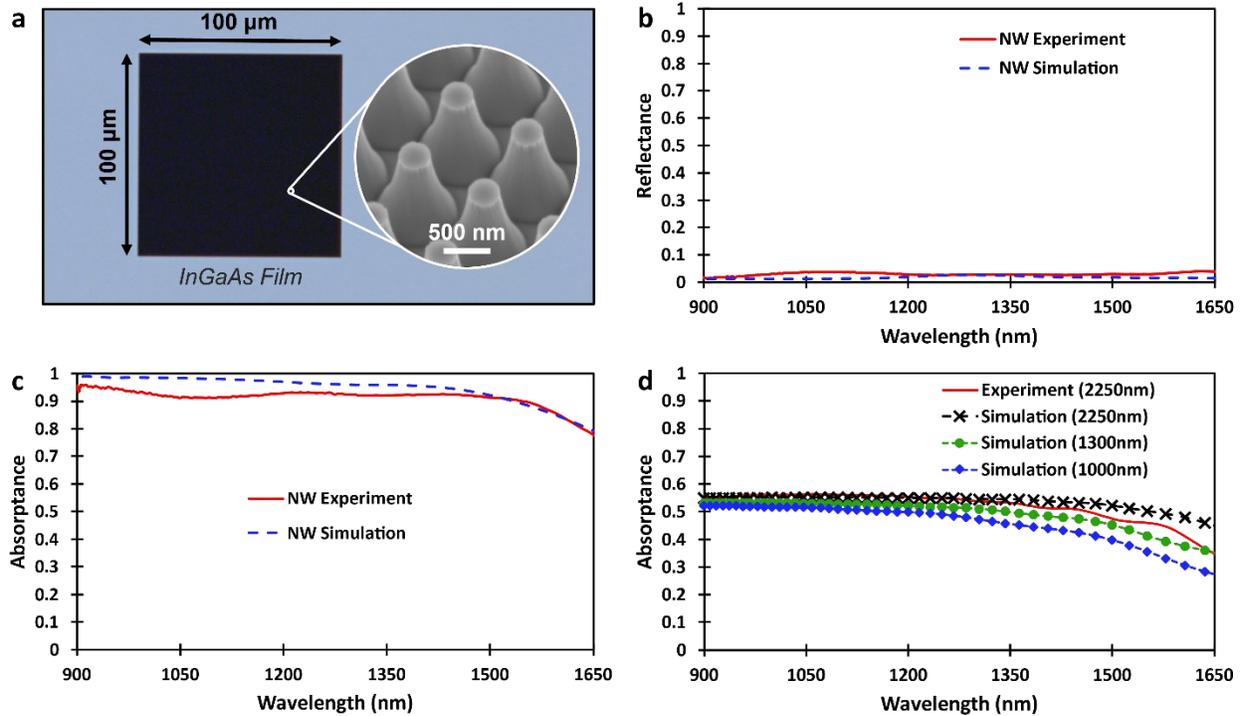

**Figure 4. Near-unity nanowire metamaterial absorption.** (a) Optical microscope image of the fabricated InGaAs tapered nanowire metamaterial (black square) on an InGaAs film (blue-grey). The lack of contrast in the 100 μm x 100 μm active area is indicative of high absorption. The inset shows a scanning electron micrograph of tapered InGaAs nanowires that is representative of the selected area from the high absorbing metamaterial region. (b) Comparison of the simulated and experimentally measured reflectance for the nanowire metamaterial as a function of wavelength. The fabricated (modelled) nanowire metamaterial dimensions are similar with a top radius: 175 nm (170 nm), bottom radius: 440 nm (441 nm), height: 1300 nm (1400 nm), and a pitch: 900 nm (900 nm). (c) Comparison of the simulated and experimentally measured absorption spectra of an InGaAs tapered nanowire metamaterial with the same dimensions as (b), which demonstrates the near-unity absorption over an unprecedented wavelength range. The measured (simulated) average absorption efficiency from 900 nm to 1500 nm is 93% (97%). (d) Measured and calculated absorption spectra of a planar InGaAs film on an InP substrate for various thicknesses - Measured film: 2250 nm; Modelled films: 1000 nm, 1300 nm, and 2250 nm.

**Conclusion**

We studied the theoretical mechanisms that govern the enhanced absorption in nanowire metamaterials, with emphasis in the infrared. We identified two prominent mechanisms that contribute to near-unity absorption in nanowire metamaterials. These include the free-space coupling of photons into the nanowires which excite leaky guided modes, followed by the strong interaction between neighbouring nanowires which excite transverse modes. We subsequently showed the importance of nanowire periodicity on achieving the maximum absorption efficiency of the metamaterial by manipulating the interaction between neighbouring nanowires. We optimized for near-unity absorption in cylindrical and tapered nanowire metamaterials, demonstrating a systematic approach to achieve narrowband and broadband characteristics. Furthermore, we fabricated a tapered InGaAs nanowire metamaterial with dimensions similar to the optimized structure. With this optimized structure, we experimentally demonstrated the near-unity absorption efficiency over an unprecedented wavelength range from 900 nm to 1500 nm, with an average efficiency of 93%. Although we demonstrated near-unity broadband absorption of the nanowire metamaterial in the near-infrared region, our approach is not limited to only InGaAs. The nanowire metamaterial can be tailored to achieve enhanced absorption in the mid-infrared and other spectral regions through the incorporation of appropriate semiconductor materials. Realizing this approach to achieve near-unity absorption efficiency over such a broad bandwidth for the first time will play a major role in the discovery of a new generation of photodetectors targeting a wider range of applications from biomedical to quantum information.

**Acknowledgement**

This research was undertaken thanks in part to funding from the Canada First Research Excellence Fund, NSERC and Industry Canada.

**Data Availability**

The data that support the plots within this paper and other findings of this study are available from the corresponding author upon reasonable request.

# Semiconductor nanowire metamaterial for broadband near-unity absorption

[†]Burak Tekcan[1,2], [†]Brad van Kasteren[1,2], [†]Sasan V. Grayli[1,2], Daozhi Shen[1,4,5], Man Chun Tam[2,3], Dayan Ban[2,3], Zbigniew Wasilewski[2,3], Adam W. Tsen[1,3,4,6], Michael E. Reimer[1,2,6]*

[1]Institute for Quantum Computing, University of Waterloo, Ontario, Canada. [2]Department of Electrical & Computer Engineering, University of Waterloo, Ontario, Canada. [3]Waterloo Institute for Nanotechnology, University of Waterloo, Ontario, Canada. [4]Department of Chemistry, University of Waterloo, Ontario, Canada. [5]Centre for Advanced Materials Joining, University of Waterloo, Ontario, Canada, [6]Department of Physics and Astronomy, University of Waterloo, Ontario, Canada.
†These authors contributed equally to this work.

## SUPPLEMENTARY 1: General Numerical Calculation Implementations

The modeling of the nanowire metamaterial devices that are presented in the manuscript are all done using the Finite-Difference Time-Domain (FDTD) Lumerical Solutions. The optical absorption characteristics of the structures were calculated using the advanced absorption analysis group provided by the modeling tool. The models were created with a fine mesh size relative to the structures (0.025 nm). For non-repeating structures, a Total-Field Scattered-Field (TFSF) source was used throughout the optimization process with perfect matched layer (PML) boundary conditions. The numerical calculations for the nanowire metamaterial arrays were run with a planewave source and Periodic Boundary Conditions (PBC).

## SUPPLEMENTARY 2: Tapered Nanowire Dimension Optimization Process

The dimensions of the tapered nanowires were optimized by first finding a cylindrical nanowire diameter that supports a single leaky resonant mode at the energy close to the bandgap of InGaAs at room temperature ($E_g \approx 0.75$ eV, $\lambda \approx 1650$ nm). FDTD Lumerical solutions was used to model the absorption characteristics of the tapered nanowire by systematically increasing the bottom diameter of the cylinder until the highest performance broadband absorption over the target spectral region was achieved. Then, the optimized tapered nanowire was placed in a periodic lattice structure and the spacing between the nanowires (periodicity) was swept over the target range of 900 nm to 1800 nm. The absorption efficiency as a function of wavelength vs. periodicity was plotted and the lattice spacing with highest broadband absorption for the target spectral range was selected.

The initial optimization was done with air as the background medium (n=1) and without a substrate. In this way, the nanowire array was suspended in air and remained the only contributor to the targeted absorption being maximized. The height of the nanowires was set to 1400 nm based on insights from previous numerical calculations of the nanowire metamaterial.

## SUPPLEMENTARY 3: Improved Model Accuracy of the Metamaterial with Tapered Nanowire Metaatoms

In order to verify the agreement between the experimentally measured absorption spectra and the calculated results from simulation, the optical absorption characteristics were obtained by measuring the absorption spectra of InGaAs nanowires fabricated on InGaAs/InP substrate. The incoming light passes through a Schwarzschild reflective objective with a numerical aperture of 0.5, introducing incident angles from -30° to 30° of the focused beam. This was accounted for in the numerical calculations by sweeping the source angle of incidence over a matching range and averaging the absorptance.

## SUPPLEMENTARY 4: Effect of Tapered Nanowire Height on the Metamaterial Absorption Efficiency

We investigated the effect of the metaatom nanowire height on the absorption efficiency by systematically reducing the nanowire height from 1.6 µm to 200 nm (Figure S1a). The lattice structure was constructed from the tapered nanowires infinitely repeating with periodicity of 900 nm without a substrate. The top and bottom diameters match the fabricated metamaterial absorbers, i.e., the top diameter is 350 nm and bottom diameter is 880 nm. It was observed that the high-performance broadband absorption efficiency over the target spectrum (0.9 µm to 1.7 µm) remains relatively unchanged for nanowire heights of 1.2 µm to 1.6 µm. As the height of the nanowires are reduced below 1.2 µm, the absorption efficiency performance is shown to noticeably decrease.

Next, the effect of the nanowire height was studied after including a substrate in the numerical calculation to reflect a more realistic comparison between simulations and the fabricated nanowire arrays. The substrate consisted of an InP wafer and a thin film of InGaAs to match the parameters of the fabricated metamaterials. The absorption response of the updated metamaterial was then found as a function of nanowire height over the high broadband performance height range found for the suspended nanowires (1 µm to 1.6 µm). The thickness of the total InGaAs (including the nanowire and substrate thickness) film remained constant, i.e., the substrate thickness was increased to maintain a total thickness of 2250 nm of the InGaAs film when the height of the nanowire was decreased. As seen in Figure S1b, the absorption efficiency showed little dependence on the changing nanowire height when the substrate is included, highlighting the impact of the substrate on the overall metamaterial absorption efficiency characteristics.

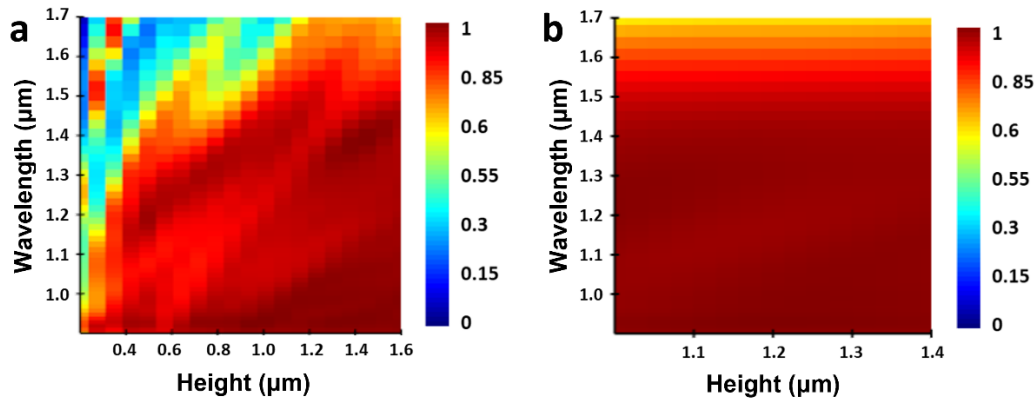

**Figure S4. The effect of height on the absorption response of the InGaAs tapered nanowire metamaterial.** The height absorption response is shown without a substrate in a) and with a substrate in b). Without the substrate, the impact of the changing metaatom height becomes apparent in the absorption efficiency of the metamaterial. Once the nanowire height is reduced below 1 µm, the high broadband absorption performance of the metamaterial begins to diminish. b) With the substrate, the broadband absorption efficiency of the metamaterial does not undergo a noticeable change with the varying nanowire height and when compared to a).

## SUPPLEMENTARY 5: Growth and Characterization of the Indium Gallium Arsenide Film

The InGaAs film that was used to fabricate the metamaterial array was grown by a Veecco GEN10 molecular beam epitaxy system, located at the Quantum Nano Centre Molecular Beam Epitaxy Lab (University of Waterloo), on a (100) InP single crystal wafer. Ellipsometry was used to extract the complex refractive index of the InGaAs film and is plotted in Figure S2. This optical data was imported into the Lumerical FDTD software and it was further used in all of the InGaAs nanowire metamaterial modeling.

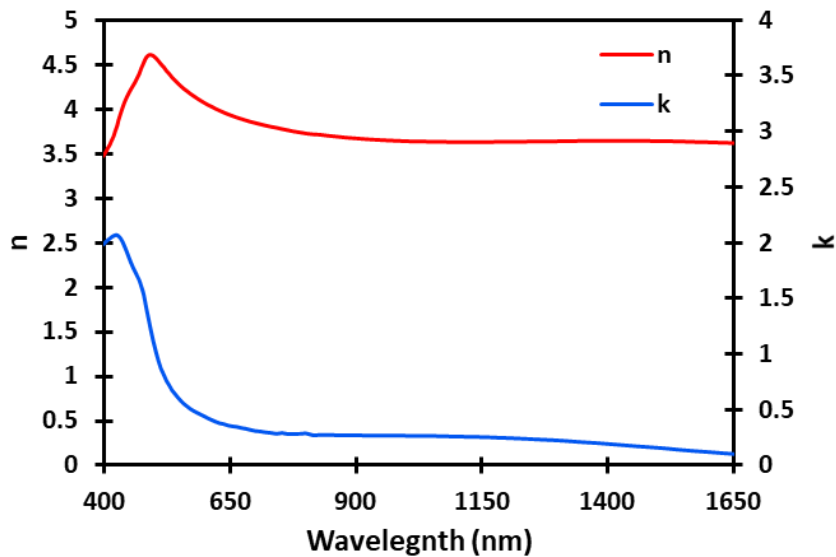

**Figure S5. Complex refractive index of InGaAs.** The spectroscopy ellipsometry measured complex refractive index of the epitaxially grown InGaAs film on a (100) InP single crystal wafer.

## SUPPLEMENTARY 6: Fabrication of the InGaAs Tapered Nanowire Metamaterial

The tapered nanowire arrays fabrication process started with the growth of a planar InGaAs epitaxial layer on a commercial InP substrate by Molecular Beam Epitaxy (MBE). The planar InGaAs thin films were cleaved into 0.75 x 0.75 cm square chips and cleaned by Acetone and IPA 5 minutes each in succession in ultrasonic tanks. The cleaned chips were loaded in a Plasma Enhanced Chemical Vapor Deposition (PECVD) system for SiNx deposition. A 500 nm thick SiNx film was deposited as a hard mask for the nanowire etch stage. Next, the chips were spun with PMMA A3 (~150nm), followed by electron beam lithography to initialize the nanowire array patterning. The chips were developed with MIBK:IPA (1:3) for 60 seconds before depositing 15 nm Al:40 nm Cr. The Al:Cr metal layer was used to transfer the e-beam lithography pattern onto the SiNx hard mask. Then, the SiNx hard mask array pattern was created by etching SiNx in an Inductively Coupled Plasma Reactive Ion Etching (ICP-RIE) system to achieve a high aspect ratio hard mask pattern. Next, the InGaAs layers were etched down to a desired depth using the ICP-RIE system. Finally, the InGaAs etched samples were dipped into a 10:1 Buffered Oxide Etchant (BOE) solution for 7 minutes to remove the remaining SiNx hard mask caps. Figure S4a illustrates the fabrication process flow of the tapered InGaAs nanowire array broadband absorber. Figure S4b and S4c show scanning electron micrographs of the fabricated InGaAs nanowire arrays.

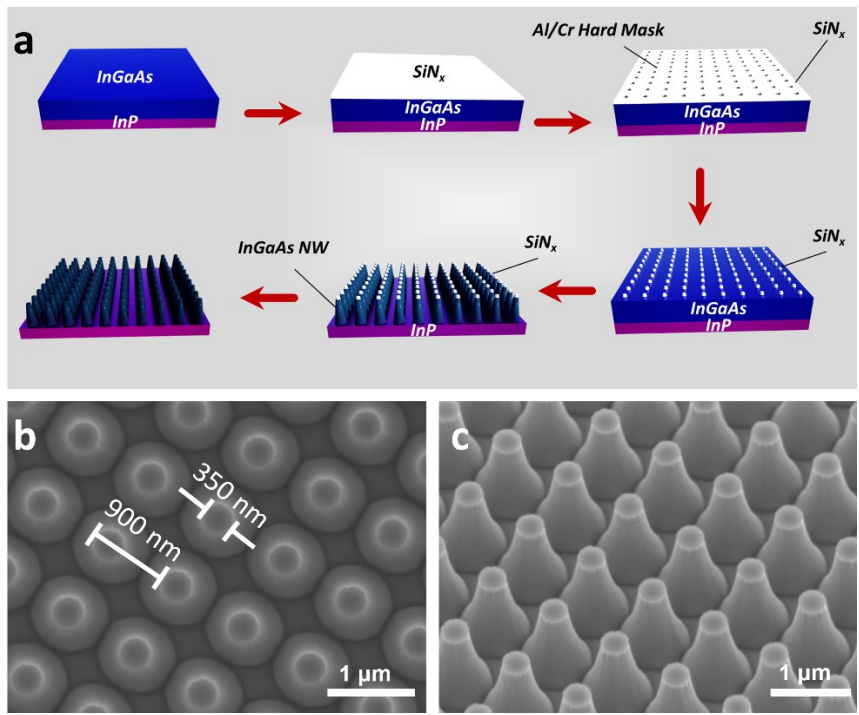

**Figure S6. Fabrication of the nanowire metamaterial.** The fabrication steps of the InGaAs tapered nanowire array are shown in a). b) and c) are the scanning electron micrograph of the fabricated nanowires with estimated periodicity of 900 nm and the top diameter of 350 nm.

## SUPPLEMENTARY 7: FTIR Measurement

The transmission and reflection spectra of the samples were measured using a Bruker IFS 66v/S Fourier-transform infrared (FTIR) spectrometer. A tungsten halogen source (OSRAM, 275W) was focused onto the sample under test by a 36× reflective objective (Ealing, NA = 0.5), following the FTIR beam splitter. For the transmission spectrum measurement, the transmitted light from sample and background was collected by another reflective objective (behind the sample) and directed to a liquid-nitrogen-cooled mercury cadmium telluride (MCT) detector (EG&G JUDSON J16D-M204-R05M-60). For the reflection measurement, a D-shaped mirror (Thorlabs PFD10-03-P01) was placed before the Schwarzschild reflective objective. The reflected light from sample was collected into the MCT detector by the D shape mirror. The D-shaped mirror used in this reflection measurement was a silver mirror (Thorlabs PF10-03-P01) with a reflectivity > 97%, this was used as the reference spectrum for the wavelength range of interest.

## SUPPLEMENTARY 8: FTIR Data Processing

The measured reflection/transmission spectra were used to extract the absorption characteristics of the fabricated metamaterial samples. Reference data was collected for the source, absent of any absorber samples. This reference data was used to normalize the FTIR data collected from the sample measurements to the source. The sample absorption efficiency was then calculated from the normalized FTIR sample data, knowing A = 1 – T – R. Where A is the absorptance, T is the transmittance, and R is the reflectance. The resulting absorption data was then adjusted to account for the artificial absorption observed below the InGaAs bandgap energy, which resulted from the thin film interference in the transmission data that was introduced from the reflective objective. To account for this interference, the data was scaled to be considerate of the InGaAs bandgap where the lowest absorption is expected to be near zero. The background correction method used scaled the absorption data according to the formula $(x_i - x_{min})/(1-x_{min})$. The results from scaling each data point by this formula bounded the lower absorptance to near zero and underestimates the absorption efficiency of the nanowire metamaterial. Despite the conservative approach, the resulting metamaterial absorption data presented an average of 93% absorption efficiency between 900 nm-1500 nm range. The absorption measurement was plotted before and after the correction in Figure S2. Here, the arrow indicates the region below bandgap where the scaled absorptance was lower bounded to near zero.

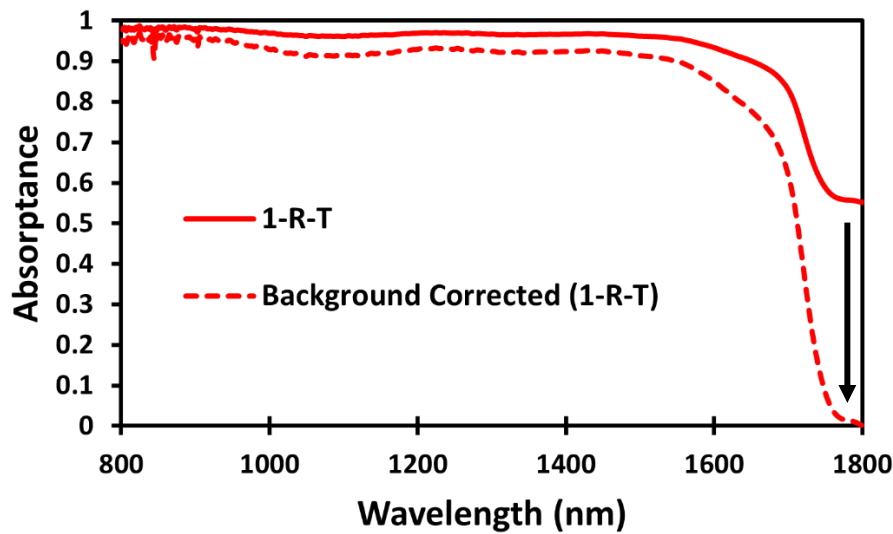

**Figure S7. The calculated absorption efficiency of the fabricated nanowire metamaterial before and after the thin film interference background correction**. This plot was obtained from subtracting the source normalized sample reflection and transmission data from unity. The arrow indicates the region below the InGaAs bandgap which was expected to exhibit a near zero absorption.

## SUPPLEMENTARY 9: Planar InGaAs FTIR Measurement Process

The optical absorption characteristics of a planar unprocessed InGaAs film (thickness=2250 nm) was obtained according to the FTIR measurement process described in the "FTIR Measurement" section. The experimental setup for the planar measurements differed slightly, however. The reflection and transmission data for the planar samples were collected without the use of a reflective objective; the planar sample was illuminated with a semi-collimated beam instead to prevent the uncertainty that needed to be adjusted for in the data observed in the nanowire sample measurements (see section "FTIR Data Processing"). The planar absorption measurement was compared with the modeled absorption response of 2250 nm, 1400 nm and 1000 nm thick InGaAs films of manuscript Figure 4d.

# SUPPLEMENTARY 10: Optical Properties of the Tapered Nanowire Metamaterials with Varying Metaatom Unit Cells

To study the success of the optimization process and simulation model, additional nanowire metamaterial designs were numerically and experimentally analyzed. This process also acted to further improve understanding for how the physical properties of the metaatoms can impact the metamaterial broadband absorption behaviour. Specifically, a series of tapered nanowire metamaterials with varying pitch and radii were investigated. The scanning electron micrographs of the various nanowire metamaterials, corresponding to NW1-NW4, are included in Figure S5. The broadband absorption responses of these various metamaterials were also experimentally assessed with FTIR, according to the "FTIR Measurement" and "FTIR Data Processing" sections. Figure S6 shows the corresponding modeled and experimentally obtained broadband absorption response of the fabricated nanowire metamaterials NW1 – NW4. We note that the metaatom nanowire height for NW1-NW4 was approximately 1.3 µm.

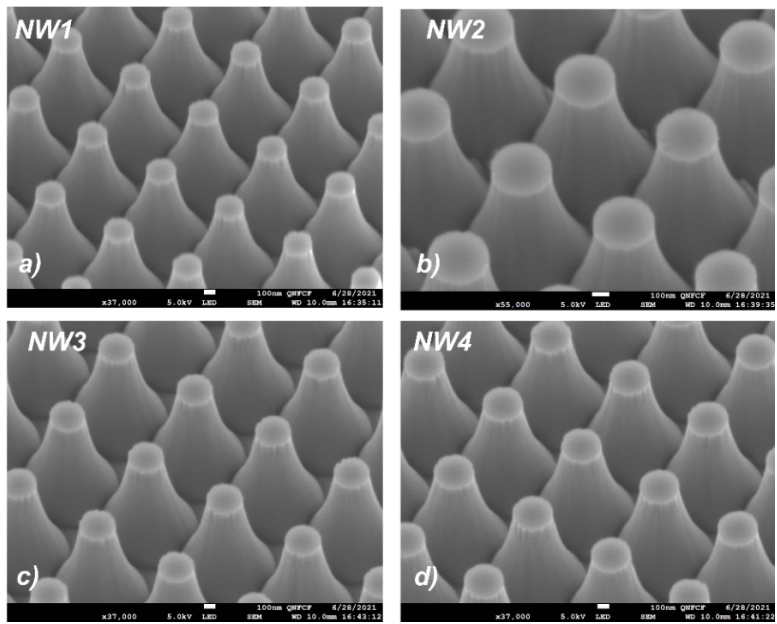

**Figure S5. The tilted angle scanning electron micrographs of the fabricated metamaterials NW1-NW4.** a) Sample NW1 was measured to have: d_top = 250 nm, d_bottom = 682 nm, periodicity = 700 nm. b) Sample NW2: d_top = 350 nm, d_bottom = 682 nm, periodicity = 700nm. c) Sample NW3: d_top = 300nm, d_bottom = 782nm, periodicity = 800nm. d) Sample NW4: d_top = 350 nm, d_bottom_782 nm, periodicity = 800 nm. The height of the nanowires in all of the fabricated metamaterials was measured to be approximately 1.3 µm.

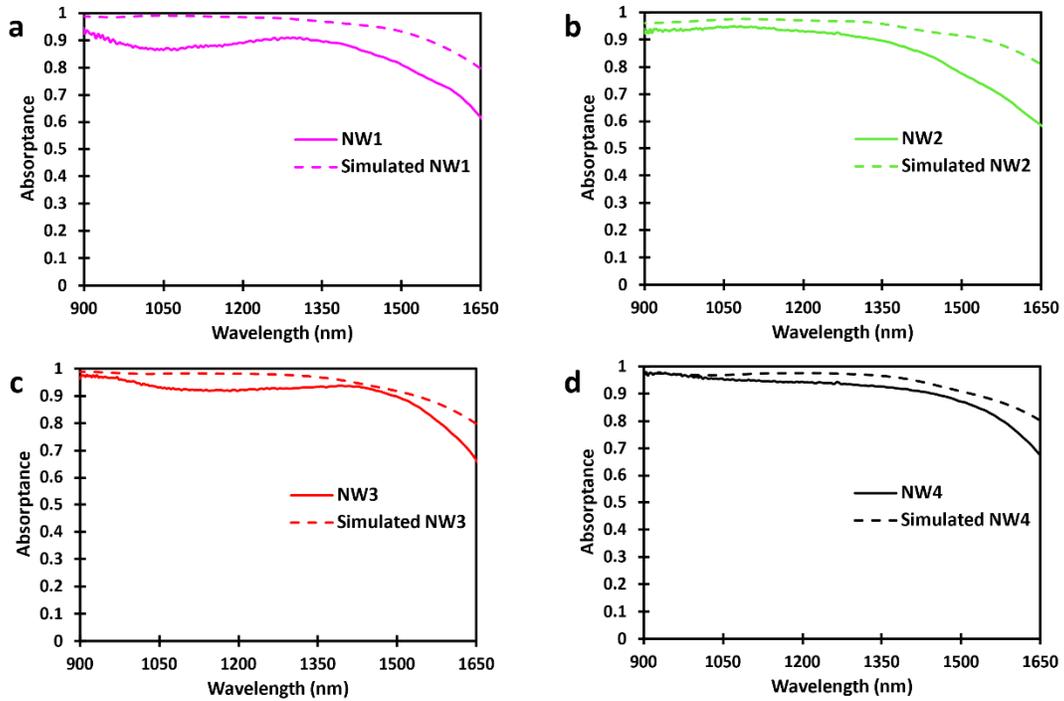

**Figure S6. The numerical and FTIR experimental absorption efficiency of four different broadband nanowire metamaterial absorbers, NW1-NW4.** a)-d) show the measured and modeled absorption efficiencies of the respective metamaterials outlined in Figure S5.